\documentclass[aps,prx,onecolumn,superscriptaddress,notitlepage,10pt,longbibliography]{revtex4-2}
\usepackage{amsthm}
\usepackage{amsmath}
\usepackage{amssymb}
\usepackage{mathrsfs}

\usepackage[only,mapsfrom]{stmaryrd}
\usepackage{enumerate}
\usepackage{todonotes,soul}
\usepackage{xcolor}
\usepackage[dvipsnames]{xcolor}
\usepackage{hyperref}
\numberwithin{equation}{section}
\newcommand{\tr}{\operatorname{tr}}

\newcommand{\id}{\mathrm{id}}
\newcommand{\ip}[2]{(#1,#2)}
\newcommand{\ii}{\mathrm{i}}
\newcommand{\ee}{\mathrm{e}}
\newcommand{\dvol}{{\rm dvol}}
\newcommand{\dd}{\mathrm{d}}
\newcommand{\1}{1\!\!1}

\newcommand{\RR}{\mathbb{R}}
\newcommand{\CC}{\mathbb{C}}
\newcommand{\NN}{\mathbb{N}}

\newcommand{\BB}{\mathcal{B}}
\newcommand{\Rf}{\mathscr{R}}
\newcommand{\Sf}{\mathscr{S}}
\newcommand{\Ac}{\mathcal{A}}
\newcommand{\HH}{\mathcal{H}}
\newcommand{\KK}{\mathcal{K}}
\newcommand{\UU}{\mathcal{U}}
\newcommand{\Y}{\yen}
\newcommand{\ac}{\mathrm{ac}}
\newcommand{\sing}{\mathrm{sing}}
\newcommand{\bra}[1]{\langle #1|}
\DeclareMathOperator{\Bor}{Bor}
\DeclareMathOperator{\supp}{supp}
\DeclareMathOperator{\ran}{ran}
\newtheorem{thm}{Theorem}[section]
\newtheorem{lemma}[thm]{Lemma}
\newtheorem{cor}[thm]{Corollary}
\newtheorem{prop}[thm]{Proposition}
\newtheorem{rem}[thm]{Remark}
\theoremstyle{definition}
\newtheorem{defn}[thm]{Definition}
\let\Re\relax
\DeclareMathOperator{\Re}{Re}
\begin{document}
\title{Quantum Reference Frames, Locality, and Microcausality in Quantum Field Theory}
\author{Christopher J. Fewster}
\email{chris.fewster@york.ac.uk}
\affiliation{
Department of Mathematics, University of York, Heslington, York YO10 5DD, United Kingdom.}
\affiliation{York Centre for Quantum Technologies, University of York, Heslington, York YO10 5DD, United Kingdom.}
\author{Leon  Loveridge}
\email{leon.d.loveridge@usn.no }
\affiliation{Quantum Technology Group, Department of Science and Industry Systems, University of South-Eastern Norway, 3616 Kongsberg, Norway}

\begin{abstract}
As a consequence of the relativistic spectrum condition,
quantum reference frames for translations on Minkowski space are
never described by projection-valued measures, and in many cases
fail a weaker spacetime localisation property known as the norm-$1$ condition. This has several implications for the local and observable structure of quantum field theories described relative to even the best available frame: for a large class of frame states, strict locality is replaced by almost locality, and microcausality by commutators that decay rapidly with increasing separation, corresponding to an effective smearing of the light
cone by the localisation density of the frame state. Operations built from finitely many almost local Kraus operators, for example nonselective measurements of relativised local observables, satisfy an
approximate no-signalling bound, with a residue attributable to the
localisation uncertainty of the frame.  Whether causality in its strongest operational form -- non-disturbance under all physically realisable operations -- survives relativisation is left open. An upshot of our finding is that quantum field theories formulated relative to a quantum reference frame cannot rest on the traditional axioms, and some general relational axiomatisation must replace at least some of them by taking into account physically realistic properties of the frame.
\end{abstract}
\maketitle

\section{Introduction}

If a physical system possesses a symmetry, every preparation procedure is accompanied by an orbit of others, related by the symmetry. The same applies to every measurement procedure.
Specifying the nature of the procedure only specifies the orbit, not a particular instance within it. Without a way to break the symmetry and identify which preparation procedure on the orbit was used, the experimenter is restricted to those measurement procedures that are invariant with respect to the symmetry. In other words, physical observables are invariants.

One way to extend the range of available observables is to introduce a \emph{reference frame} - a physical system that shares the same symmetry as the original, for then the invariant observables of the joint system become available. For example, while the position of a single particle in Euclidean space is not translationally invariant, the relative position of the particle with respect to a reference particle is invariant under common translations of both particles. If the reference particle is in a sufficiently well localised state, the relative position behaves, to good approximation, like the position of the original particle; the relative position is the translation invariant object that is measured in practice, with the absolute position being a good proxy in the limit of a perfectly localised reference.

Let us describe how this programme is implemented within quantum theory. Here we consider
a quantum system $\Sf$, and a reference $\Rf$, itself a quantum system. For this reason, $\Rf$ is called a quantum reference frame (QRF). We suppose that both $\Sf$ and $\Rf$ admit a symmetry described by a symmetry group $G$, which is assumed to be locally compact. The mathematical description of $\Sf$ and $\Rf$, and of $\Sf+\Rf$, includes quantities that would be observable in the absence of the symmetry. For want of a better term, we temporarily describe these as `potential observables'. If $\Sf$ is a quantum-mechanical system, one may always find an \emph{ideal} QRF $\Rf$ so that the algebra of potential observables of $\Sf$ is isomorphic to an algebra of invariant \emph{relative observables} of $\Sf + \Rf$ (e.g. \cite{bartlett2007reference,loveridge2012quantum,loveridge2018symmetry,de2021perspective,glowacki2024quantum,carette2025operational}), and this algebra can be recovered arbitrarily precisely via a physically well motivated conditional expectation, which gives the description of the system contingent on the specification of the state of the frame. This indicates that for quantum mechanical systems, it is to some degree a matter of taste
whether to include the frame in the description, or not: in this sense, all potential observables become truly observable. Reference frames that are described by quantum mechanical systems are now well studied (e.g., \cite{bartlett2007reference,giacomini2019quantum,vanrietvelde2020change,de2021perspective,carette2025operational,de2020quantum,castro2025relative}).

The situation changes when considering quantum field theory, where the axiomatic basis for the theory puts more stringent conditions on the frame; here the ideality property may be lost, and the above-mentioned isomorphism is typically no longer tenable.
More precisely, given a locally compact group $G$ with a strongly continuous unitary representation $U_{\Rf}$ in some Hilbert space $\HH_{\Rf}$,  a (principal) quantum reference frame is defined to be a \emph{system of covariance} based on $G$, that is, the triple $(\HH_{\Rf},U_{\Rf},E)$, where $E$ is a (possibly subnormalised) positive operator valued measure satisfying the covariance condition
\begin{equation}\label{eq:covpovm1}
    U(g)E(\Delta)U(g)^* = E(g.\Delta)~~\text{for all}~~g \in G, \Delta \in \Bor(G).
\end{equation}
From here, the \emph{relativisation map} (e.g. \cite{loveridge2012quantum,loveridge2018symmetry}) $\Y: \BB(\HH_{\Sf}) \to \BB(\HH_{\Sf}\otimes \HH_{\Rf})$  is given as
\begin{equation}
    \Y(A) = \int_G U(g)AU(g)^* \otimes E(dg),
\end{equation}
where the integral is understood in an appropriate manner (\cite{loveridge2018symmetry,Jamesproj,glowacki2026mathrm}).

Relativisation is understood as giving a description of $\Sf$ relative to the frame $E$; $\Y$ is a normal completely positive subunital map whose image lies in the invariant algebra $\BB(\HH_{\Sf}\otimes \HH_{\Rf})^G$, and an injective $^*$-homomorphism if $E$ is projection-valued. Those operators in $\BB(\HH_{\Sf})$ not invariant under $G$ are understood as not representing bona fide physical quantities, and acquire their physical meaning only relative to an appropriate frame.
Given that one of the motivations for introducing quantum reference frames is based on the view that the frame should be described by the same theory as the system of interest, it is important to construct quantum reference frames appropriate to quantum field theory, ideally that are themselves QFTs.

In this paper, we study quantum reference frames in Minkowski space that are covariant under ($4$-)translation, hence with $G=\RR^4$. It is known (due to a result of Toller, \cite{toller1999localization}; see also \cite{mazzucchi2001observables}) that no
POVM satisfying the covariance requirement \eqref{eq:covpovm1} plus the spectrum condition can be projection-valued, and therefore that multiplication is not preserved by $\yen$ in general, and some extra `fuzziness' is present. Moreover, these POVMs typically do not satisfy a weaker `localisability' condition, known as the norm-$1$ property, which allows the system algebra to be recovered up to arbitrarily good approximation under a natural conditional expectation. Our assumptions are weaker than those of Wightman/AQFT in Minkowski, so although our QRFs are not constructed from systems of fields, they would certainly apply  in that setting.

Our main findings are as follows. Under the spectrum condition, no
translation-covariant frame is projection-valued (Corollary \ref{cor:nopvm}, following Toller \cite{toller1999localization}), and a quantitative bound rules out even the norm-$1$ property for a large class of frames (Theorem \ref{thm:norm1}). Relative to a frame whose
localisation density is a Schwartz function, relativised local observables are \emph{almost local} (Proposition \ref{cor:al}; see \cite{haag1996,dybalski2014criterion}), their commutators at causal separation $L$ are $O(L^{-\infty})$ (Lemma \ref{lem:aloccomm}), and the sharp light cone is
replaced by its smearing with the autocorrelation of the localisation density (Remark \ref{rem:smearedcone}). Causality nonetheless  survives in some form: relativisation preserves compatibility (Proposition \ref{prop:cpcompat}), and operations with finitely many almost local Kraus operators, for instance the L\"uders measurements of relativised local observables, disturb causally distant relativised observables only at order $O(L^{-\infty})$ (Proposition \ref{prop:nosignal}, Corollary \ref{cor:relnosignal}), locality
of operations being, under Haag duality, equivalent to non-signalling (Remark \ref{rem:rigidity}). Finally, accurate recovery of local observables requires the frame to resolve small translations, and hence to carry correspondingly high energies (Proposition \ref{prop:tradeoff}).

A complementary, axiomatic development has recently been given by Fedida and G\l{}owacki \cite{fedida2025foundations}, who formulate relational covariance and causality conditions for Poincar\'e-covariant frame observables. Their causality results are built on frame preparations with compactly supported localisation distributions; however, as follows from Proposition \ref{prop:toller1} below, no such preparations exist once the frame itself satisfies the spectrum condition usually assumed in QFT \cite{streater2000pct,haag1996}. By contrast, in this work we enforce the spectrum condition at the expense of microcausality but, as we shall see, this does not necessarily entail acausal signalling.

\section{Quantum Reference Frames for Translations in Minkowski space}
Here we present Toller's construction \cite{toller1999localization} of translation-covariant POVMs on (here, $1+3$-dimensional) Minkowski space $M$, adding some further particulars where they may be helpful.  Throughout, the Minkowski product is $k\cdot x = k^0x^0 - \vec{k}\cdot\vec{x}$, and Fourier transforms of vector-valued functions are normalised so that $(\mathcal{F}^{-1}F)(x)=\int_{\RR^4}\ee^{-\ii k\cdot x}F(k) \dd^4 k/(2\pi)^4$, unitary from $L^2(\RR^4,\dd^4 k/(2\pi)^4)$ to $L^2(M,\dd^4 x)$.

Let $\HH$ be a complex, separable Hilbert space carrying a strongly continuous unitary representation of translations
\begin{equation}
    U:\RR^4 \to \UU(\HH), \qquad U(x) = \ee^{\ii x_\mu P^\mu},
\end{equation}
whose generators satisfy the spectrum condition - that the joint spectrum of the $P^\mu$ is contained in ${V}^+$, the closed forward causal cone. We seek a POVM $E : \Bor(M) \to \BB(\HH)^+$ satisfying:
\begin{eqnarray} \label{eq:pov1}
    & U(x)E(\Delta)U(x)^* = E(\Delta + x), \\
    & 0  \neq E(M)\le \1_\HH.\label{eq:pov2}
\end{eqnarray}
Note that the usual normalisation $E(M)=\1_\HH$
 is too strong and is replaced by subnormalisation \eqref{eq:pov2}.
 If $E(\Delta)$ is interpreted as an effect associated with a certain event occurring in $\Delta$ (or a measurement returning a value in $\Delta$), then $E(M)$ is the effect corresponding to the event occurring at all (or that the measurement succeeds in returning a result).  As will be seen below (Theorem \ref{thm:toller}), $E$ vanishes on the singular subspace $\mathcal{H}_{\mathrm{sing}}$ of the energy-momentum spectral measure, which contains the vacuum and any single-particle states. Vectors in
$\mathcal{H}_{\mathrm{sing}}$ therefore do not define events in this sense; more generally, writing $\psi=\psi_{\mathrm{ac}}+\psi_{\mathrm{sing}}$, one has
$(\psi,E(M)\psi) =(\psi_{\mathrm{ac}},E(M)\psi_{\mathrm{ac}})\le\|\psi_{\mathrm{ac}}\|^2$, so the singular component of a state contributes only to the `non-occurrence event'.

 Furthermore, in any QFT with a translation-invariant vacuum vector that is separating for the local algebras (as guaranteed by the Reeh--Schlieder theorem), no nonzero $E(\Delta)$ can belong to a local algebra, and under the cluster property not even to the quasilocal algebra \cite{giannitrapani1998quantum}.
The next proposition collects two related facts which together are the foundation of much of what follows. Both are due to Toller \cite{toller1999localization}; for part (ii) we write the proof out in detail.

\begin{prop}\label{prop:toller1}
Let $E:\Bor(M)\to \BB(\HH)$ be a POVM satisfying \eqref{eq:pov1}--\eqref{eq:pov2}.
\begin{enumerate}
\item[(i)] For every nonempty open set $\Delta\subset M$,
the positive operator $E(\Delta)$ is non-zero.
\item[(ii)] For every nonempty open set $\Delta\subset M$ and every $\psi\in\HH$,
\begin{equation}
    \ip{\psi}{E(\Delta)\psi} = 0 ~\Longrightarrow ~\ip{\psi}{E(M)\psi} = 0.
\end{equation}
\end{enumerate}
\end{prop}
\begin{proof}
\emph{(i)} Suppose $E(\Delta)=0$ for some nonempty open $\Delta$. Choose $\epsilon>0$ such that $\Delta$ contains an open ball $B_\epsilon$ of radius $\epsilon$. Countably many translates of $B_\epsilon$ cover $\RR^4$, and by \eqref{eq:pov1},
\begin{equation*}
    E(B_\epsilon+x) = U(x)E(B_\epsilon)U(x)^* \le U(x)E(\Delta)U(x)^* = 0.
\end{equation*}
Countable additivity of $E$ then gives $E(M)=0$, contradicting \eqref{eq:pov2}.
\emph{(ii)} Suppose $\ip{\psi}{E(\Delta)\psi}=0$ for some nonempty open $\Delta$, and choose nonempty open sets $\Delta',\Delta''$ with $\Delta'+\Delta''\subseteq\Delta$ (two balls of half the radius of a  ball contained in $\Delta$ suffice). For $x\in\Delta''$ we have $\Delta'+x\subseteq\Delta$, and therefore monotonicity of the POVM together with \eqref{eq:pov1} gives
\begin{equation}\label{eq:psiEzero}
    0\le \|E(\Delta')^{1/2}U(x)^*\psi\|^2 = \ip{\psi}{E(\Delta'+x)\psi} \le \ip{\psi}{E(\Delta)\psi} = 0,
\end{equation}
so $E(\Delta')^{1/2}U(x)^*\psi = 0$ for all $x\in\Delta''$.
Now, fix any $\varphi\in\HH$ and define
\begin{equation*}
    f(z) = \ip{E(\Delta')^{1/2}\varphi}{\ee^{-\ii z\cdot P}\psi}, \qquad z\in\CC^4.
\end{equation*}
The spectrum condition $\sigma(P)\subseteq {V}^+$ implies that $\ee^{-\ii z\cdot P}\psi$ is well defined as an $\HH$-valued holomorphic function of $z$ in the lower tube $T_- = \RR^4 + \ii(-V)$, where $V$ denotes the interior of $V^+$, with continuous boundary values on $\RR^4$ (by strong continuity of $U$ and uniform boundedness of $\|\ee^{-\ii z\cdot P}\psi\|$ on $\overline{T_-}$). Hence $f$ is holomorphic in $T_-$, continuous on the boundary, and by \eqref{eq:psiEzero} vanishing on $\Delta''$. Setting $f^*(z) = \overline{f(\bar z)}$ gives a holomorphic function on the upper tube $T_+ = \RR^4+\ii V$ whose boundary values agree with those of $f$ on $\Delta''$. The edge-of-the-wedge theorem \cite{streater2000pct} then provides a holomorphic continuation across $\Delta''$, giving a single holomorphic function on a connected open set containing $T_-\cup\Delta''\cup T_+$. As this continuation vanishes on $\Delta''$ and the domain is connected, it vanishes identically. In particular, $f(x)=0$ for every $x\in\RR^4$, and since $\varphi\in\HH$ was arbitrary, $E(\Delta')^{1/2}U(x)^*\psi=0$ for every $x\in\RR^4$. Equivalently,
\begin{equation*}
    \ip{\psi}{E(\Delta'+x)\psi} = 0 ~~\text{for all } x\in\RR^4.
\end{equation*}
Applying the covering of part (i) to $\Delta'$ now yields $\ip{\psi}{E(M)\psi}=0$.
\end{proof}
Part (ii) of Proposition \ref{prop:toller1} is enough to rule out projection-valued $E$:
\begin{cor}\label{cor:nopvm}
No POVM $E:\Bor(M)\to\BB(\HH)$ satisfying \eqref{eq:pov1}--\eqref{eq:pov2} is projection-valued.
\end{cor}
\begin{proof}
Suppose, for a contradiction, that $E$ is a PVM. For any nonempty open $\Delta\subset M$ and any $\psi\in\ker E(\Delta)$, $\ip{\psi}{E(\Delta)\psi}=0$, so by Proposition \ref{prop:toller1}(ii), $E(M)\psi=0$. Thus $\ker E(\Delta)\subseteq\ker E(M)$, equivalently $\ran E(M)\subseteq\ran E(\Delta)$. Monotonicity gives the reverse inclusion. As both are projections (since $E$ is assumed PVM), $E(\Delta)=E(M)$ for every nonempty open $\Delta$. Now choose any bounded open $\Delta$; then $M\setminus\overline{\Delta}$ is open and nonempty, so Proposition \ref{prop:toller1}(i) gives $E(M\setminus\overline{\Delta})\neq 0$. But $E(M\setminus\overline{\Delta}) = E(M) - E(\overline{\Delta}) \le E(M) - E(\Delta) = 0$, a contradiction.
\end{proof}
We now turn to the structure of translation-covariant POVMs. The spectral theorem for the commuting self-adjoint operators $P^\mu$ provides a direct integral decomposition
\begin{equation}\label{eq:directintegral}
    \HH \cong \int^\oplus_{\RR^4} \dd\mu(k) \HH(k),
\end{equation}
where $\mu$ is supported in $V^+$. The decomposition $\mu = \mu_\ac + \mu_\sing$ induces a corresponding orthogonal decomposition $\HH = \HH_\ac \oplus \HH_\sing$; the vacuum and any single-particle sectors are contained in $\HH_\sing$, on which $E(M)$ vanishes.
After rescaling the inner products on $\HH(k)$ if necessary, the absolutely continuous part can be written as
\begin{equation*}
    \HH_\ac = \int^\oplus_{V_\ac^+} \frac{\dd^4 k}{(2\pi)^4} \HH(k),
\end{equation*}
where $V_\ac^+ = \supp\mu_\ac$.
In a theory with mass gap, $\sigma(P_\mu P^\mu) \subseteq \{0\}\cup[m^2,\infty)$ with $m>0$, one has $V_\ac^+\subseteq\{k\in V^+: k_\mu k^\mu \ge 4m^2\}$.
\begin{lemma}[Toller, {\cite[Prop. 2]{toller1999localization}}]\label{lem:smoothdensity}
For integers $r\ge 0$ write $\|\psi\|_r := \|(1+P^0)^r\psi\|_\HH$, and let $D:=\bigcap_{r\ge 0}D\bigl((1+P^0)^r\bigr)$, equipped with the Fr\'echet topology generated by the seminorms $\{\|\cdot\|_r\}_{r=0,1,2,\ldots}$; $D$ is dense and translation-invariant in $\HH$. For every $\varphi,\psi\in D$, the complex measure $\Delta\longmapsto\ip{\varphi}{E(\Delta)\psi}$ is absolutely continuous with respect to Lebesgue measure on $M$, and is represented by a smooth density $\rho_{\varphi,\psi}\in C^\infty(M)$:
\begin{equation}\label{eq:lem_smoothdensity}
    \ip{\varphi}{E(\Delta)\psi} = \int_\Delta \rho_{\varphi,\psi}(x)\dd^4 x \qquad \text{for all Borel } \Delta\subseteq M.
\end{equation}
The density is sesquilinear in $(\varphi,\psi)$, positive in the sense that $\rho_{\psi,\psi}(x)\ge 0$, and satisfies the covariance relation
\begin{equation}\label{eq:lem_covariance}
    \rho_{U(y)\varphi,U(y)\psi}(x+y) = \rho_{\varphi,\psi}(x), \qquad x,y\in M,
\end{equation}
together with the bound
\begin{equation}\label{eq:lem_pointwise}
    |\rho_{\varphi,\psi}(x)| \le 2^4\|\varphi\|_4\|\psi\|_4, \qquad x\in M.
\end{equation}
In particular, $\int_M \rho_{\psi,\psi}(x)\dd^4 x = \ip{\psi}{E(M)\psi} \le \|\psi\|^2$.
\end{lemma}

The next theorem gives a basis-free formulation of Toller's representation of the density $\rho_{\varphi,\psi}$ of translation-covariant POVMs, which can be cast as a Naimark dilation.
\begin{thm}\label{thm:toller}
Let $E:\Bor(M)\to\BB(\HH)$ be a POVM satisfying \eqref{eq:pov1}--\eqref{eq:pov2}. Then $E$ vanishes on $\HH_\sing$, and there exist a Hilbert space $\widetilde{\HH}$ and a measurable family of contractions
\begin{equation*}
    K(k):\HH(k)\to\widetilde{\HH}, \qquad \|K(k)\|\le 1 \text{ for a.e. } k\in V_\ac^+,
\end{equation*}
such that the $L^2$ map
\begin{equation}\label{eq:Wdef}
    W\psi = \mathcal{F}^{-1}(K(\cdot)\psi(\cdot))
\end{equation}
defines a contraction $W:\HH_\ac\to L^2(M,\widetilde{\HH})$, and the POVM $E$ is the (subnormalised, since $W$ is a contraction) compression of the canonical dilation on $L^2(M;\widetilde{\HH})$:
\begin{equation}\label{eq:Erep}
    E(\Delta) = W^*M_{\chi_\Delta}W, \qquad \Delta\in\Bor(M),
\end{equation}
where $M_{\chi_\Delta}$ denotes multiplication by the indicator function $\chi_\Delta$ on $L^2(M;\widetilde{\HH})$.  Equivalently,
\begin{equation}\label{eq:Erep_inner}
    \ip{\varphi}{E(\Delta)\psi} = \int_\Delta \ip{(W\varphi)(x)}{(W\psi)(x)}_{\widetilde{\HH}}\dd^4 x
\end{equation}
for all $\varphi,\psi\in\HH$ and Borel $\Delta\subseteq M$, where $W$ is extended by zero on $\HH_{\rm sing}$. Conversely, any such measurable family $\{K(k)\}$ of contractions, not vanishing almost everywhere, defines a translation-covariant POVM (vanishing on $\HH_{\rm sing}$) via \eqref{eq:Wdef}--\eqref{eq:Erep}. Moreover, $E(M)=\Pi_\ac$ (the projection onto $\HH_\ac$) if and only if almost every $K(k)$ is an isometry.
\end{thm}
\begin{proof}
Translating Toller’s component formula \cite[\S 3]{toller1999localization} into our notation, the density of any translation-covariant POVM satisfying \eqref{eq:pov1}-\eqref{eq:pov2} vanishes when either argument lies in $\HH_\sing$, and on $\HH_\ac$ takes the form
\begin{equation*}
    \rho_{\varphi,\psi}(x) = \Big\langle \int_{V_\ac^+} \ee^{-\ii k\cdot x} K(k)\varphi(k)\frac{\dd^4 k}{(2\pi)^4},\; \int_{V_\ac^+} \ee^{-\ii k\cdot x} K(k)\psi(k)\frac{\dd^4 k}{(2\pi)^4}\Big\rangle_{\widetilde{\HH}}
\end{equation*}
for some Hilbert space $\widetilde{\HH}$ and measurable field of contractions $K(k):\HH(k)\to\widetilde{\HH}$, in the usual direct-integral sense, with $\|K(k)\|\le 1$ a.e.\ on $V_\ac^+$. Defining $W:\HH_\ac\to L^2(M;\widetilde{\HH})$ by \eqref{eq:Wdef}, this reads $\rho_{\varphi,\psi}(x) = \ip{(W\varphi)(x)}{(W\psi)(x)}_{\widetilde{\HH}}$. Plancherel's theorem gives
\begin{equation*}
    \|W\psi\|^2_{L^2(M;\widetilde{\HH})} = \int_{V_\ac^+}\|K(k)\psi(k)\|^2_{\widetilde{\HH}}\frac{\dd^4 k}{(2\pi)^4} \le \int_{V_\ac^+}\|\psi(k)\|^2_{\HH(k)}\frac{\dd^4 k}{(2\pi)^4} = \|\psi\|_\HH^2,
\end{equation*}
so $W$ is a contraction, and integration of $\rho_{\varphi,\psi}$ over $\Delta$ gives \eqref{eq:Erep_inner}, equivalently the Naimark-dilation formula \eqref{eq:Erep}. In particular $E(M) = W^*W$; hence $E(M) = \Pi_\ac$ iff $\|W\psi\| = \|\psi\|$ for every $\psi \in \HH_\ac$, which by the Plancherel identity above is equivalent to $K(k)^*K(k) = \1_{\HH(k)}$ a.e., i.e., to almost every $K(k)$ being an isometry.
The converse direction - that any measurable field $K(k)$ of contractions, in the usual direct-integral sense, defines a translation-covariant POVM via \eqref{eq:Wdef}--\eqref{eq:Erep} - is a direct computation.
\end{proof}
If $\widetilde{\HH}$ is chosen minimal, in the sense that the ranges of the $K(k)$ together span a dense subspace, the pair $(\widetilde{\HH},K)$ is unique up to a unitary on $\widetilde{\HH}$.
\begin{rem}\label{rem:hpw}
On $D$, the $L^2$ equivalence class $W\psi$ has a smooth pointwise representative given by
\begin{equation*}
    (W\psi)(x) = hU(x)^*\psi, \qquad h\psi = \int_{V_\ac^+} K(k)\psi(k)\frac{\dd^4 k}{(2\pi)^4},
\end{equation*}
to be understood weakly, or as the value at $x=0$ of the Fourier transform on $D$. In general $h$ does not extend to a bounded operator $\HH_\ac\to\widetilde{\HH}$. The boundedness or otherwise of $h$ will play a structural role in the next subsection.
\end{rem}

\subsection*{Localisability}
 Theorem \ref{thm:toller} allows us to address whether translation-covariant POVMs admit the \emph{norm-$1$ property}.
\begin{defn}\label{def:norm1}
A POVM $E:\Bor(M)\to\BB(\HH)$ satisfies the \emph{norm-$1$ property} if $\|E(\Delta)\|_{\rm op} = 1$ for every nonempty open $\Delta\subseteq M$.
\end{defn}
This has two equivalent formulations (i) for every nonempty open $\Delta\subseteq M$ there is a sequence of unit vectors $(\varphi_n)\subset \HH$ for which $\lim_{n \to \infty} \ip{\varphi_n}{E(\Delta)\varphi_n}=1$ (ii) for every nonempty open $\Delta\subseteq M$ and any $\epsilon > 0$ there is a unit vector $\varphi_{\epsilon} \in \HH$ for which $\ip{\varphi_\epsilon}{E(\Delta)\varphi_\epsilon}>1-\epsilon$. The norm-$1$ property is strictly weaker than being projection-valued, but characterises those observables for which arbitrarily sharp localisation in any open region is possible. By Corollary \ref{cor:nopvm}, no translation-covariant POVM on $M$ is projection-valued, so it is natural to ask whether they may still satisfy the norm-$1$ condition.
\begin{thm}\label{thm:norm1}
Let $E:\Bor(M)\to\BB(\HH)$ be a translation-covariant POVM satisfying $0 \neq E(M)\le\1_\HH$, and let $W:\HH_\ac\to L^2(M,\widetilde{\HH})$ be the contraction as given above, so that $E(\Delta) = W^*M_{\chi_\Delta}W$. Suppose that the map
\begin{equation*}
    \psi\mapsto (W\psi)(0),
\end{equation*}
initially defined on the domain $D$, extends to a bounded operator $h:\HH_\ac\to\widetilde{\HH}$. By translation covariance, this is equivalent to asserting that the pointwise representative of $W\psi$ is given by
\begin{equation*}
    (W\psi)(x) = hU(x)^*\psi
\end{equation*}
for all $\psi\in\HH_\ac$ and $x\in M$, with $h$ bounded on $\HH_\ac$. Then, for every Borel set $\Delta\subseteq M$ of finite Lebesgue measure,
\begin{equation}\label{eq:norm1bound}
    \|E(\Delta)\|_{\rm op} \le \|h\|^2_{\rm op}|\Delta|.
\end{equation}
Consequently,
\begin{equation*}
    \sup_{|\Delta|\le\varepsilon}\|E(\Delta)\|_{\rm op} \le \|h\|^2_{\rm op}\varepsilon,
\end{equation*}
and $E$ does not have the norm-$1$ property.
\end{thm}
\begin{proof}
For the asserted equivalence: given $\psi\in\HH_\ac$, choose $\psi_n\in D$ with $\psi_n\to\psi$; then $hU(\cdot)^*\psi_n\to hU(\cdot)^*\psi$ uniformly, while $W\psi_n\to W\psi$ in $L^2$, so by Remark \ref{rem:hpw} the continuous function $x\mapsto hU(x)^*\psi$ represents $W\psi$.
Now let $\psi\in\HH_\ac$ with $\|\psi\|\le 1$. By the hypothesis and unitarity of $U(x)$,
\begin{equation*}
    \|(W\psi)(x)\|^2_{\widetilde{\HH}} = \|hU(x)^*\psi\|^2_{\widetilde{\HH}} \le \|h\|^2_{\rm op}\|\psi\|^2 \le \|h\|^2_{\rm op}
\end{equation*}
for all $x\in M$. Hence
\begin{equation*}
    \ip{\psi}{E(\Delta)\psi} = \int_\Delta \|(W\psi)(x)\|^2_{\widetilde{\HH}}\dd^4 x \le \|h\|^2_{\rm op}|\Delta|.
\end{equation*}
Taking the supremum over $\|\psi\|\le 1$ (and using that $E$ vanishes on $\HH_\sing$) gives \eqref{eq:norm1bound}. If $E$ had the norm-$1$ property, every nonempty open $\Delta$ would satisfy $\|E(\Delta)\|=1$; choosing a nonempty bounded open $\Delta$ with $|\Delta|<\|h\|_{\rm op}^{-2}$ contradicts the estimate (and if $h=0$ the conclusion is immediate), so $E$ does not have the norm-1 property.
\end{proof}
A simple, explicit sufficient condition for the hypothesis of Theorem \ref{thm:norm1} is square-integrability of the Hilbert–Schmidt norm of $K$ in momentum space.
\begin{cor}\label{cor:HS_norm1}
Suppose
\begin{equation}\label{eq:HSi}
    \int_{V_\ac^+} \|K(k)\|^2_{\rm HS}\frac{\dd^4 k}{(2\pi)^4} =: C^2 < \infty.
\end{equation}
Then the evaluation map $\psi\mapsto(W\psi)(0)$ extends to a bounded operator $h:\HH_\ac\to\widetilde{\HH}$ with $\|h\|_{\rm op}\le C$. Consequently $\|E(\Delta)\|_{\rm op}\le C^2|\Delta|$ for every Borel $\Delta\subseteq M$ of finite Lebesgue measure and $E$ does not have the norm-$1$ property.
\end{cor}
\begin{proof}
Choose an orthonormal basis $\{e_n\}$ of $\widetilde{\HH}$ and let $\varphi_n\in\HH_\ac$ be the section $k\mapsto K(k)^*e_n\in\HH(k)$. By Fubini and the definition of the Hilbert-Schmidt norm, we have
\begin{equation*}
    \sum_n \|\varphi_n\|^2_\HH = \int_{V_\ac^+} \sum_n \|K(k)^*e_n\|^2_{\HH(k)}\frac{\dd^4 k}{(2\pi)^4} = \int_{V_\ac^+} \|K(k)\|^2_{\rm HS} \frac{\dd^4 k}{(2\pi)^4} = C^2.
\end{equation*}
For $\psi\in\HH_\ac$ and $x\in M$,
\begin{equation*}
    \ip{e_n}{(W\psi)(x)}_{\widetilde{\HH}} = \int_{V_\ac^+}\ee^{-\ii k\cdot x}\ip{K(k)^*e_n}{\psi(k)}_{\HH(k)} \frac{\dd^4 k}{(2\pi)^4} = \ip{\varphi_n}{\ee^{-\ii x\cdot P}\psi}_\HH,
\end{equation*}
and Cauchy-Schwarz gives $|\ip{\varphi_n}{\ee^{-\ii x\cdot P}\psi}|\le\|\varphi_n\|\|\psi\|$. Summing,
\begin{equation*}
    \|(W\psi)(x)\|^2_{\widetilde{\HH}} = \sum_n |\ip{\varphi_n}{\ee^{-\ii x\cdot P}\psi}|^2 \le \sum_n \|\varphi_n\|^2_\HH\|\psi\|^2_\HH = C^2\|\psi\|^2_\HH,
\end{equation*}
so the hypothesis of Theorem \ref{thm:norm1} holds with $\|h\|_{\rm op}\le C$.
\end{proof}
\begin{cor}\label{cor:sem1}
If $\widetilde{\HH}=\CC$ and $K(k)=\bra{\varphi(k)}$ for some $\varphi\in\HH$ with $\|\varphi(k)\|_{\HH(k)}\le 1$ a.e.\ and $\|\varphi\|_\HH\le 1$, then $(W\psi)(x)=\ip{\varphi}{\ee^{-\ii x\cdot P}\psi}$ is bounded in absolute value by $\|\psi\|_\HH$, and
\begin{equation*}
    \|E(\Delta)\|_{\rm op}\le |\Delta|
\end{equation*}
for every Borel $\Delta\subseteq M$ with $|\Delta|<\infty$. Thus, $E$ does not have the norm-$1$ property.
\end{cor}

\begin{rem}\label{rem:Tol5}
Theorem \ref{thm:norm1} and Toller's \cite[Prop. 5]{toller1999localization}
are complementary. Theorem \ref{thm:norm1} shows that boundedness of $h$, which is guaranteed, for instance, by condition \eqref{eq:HSi}, obstructs the norm-$1$ property being satisfied, with the quantitative bound \eqref{eq:norm1bound}. Toller gives a sufficient condition for the norm-$1$ property to hold, given by the asymptotic constancy of $K(k)$ as $|k|\to\infty$, under which a translation-covariant POVM is in Toller's terminology \emph{definite}, which  coincides with the norm-$1$ property. In particular, any definite translation-covariant POVM has unbounded $h$, by Theorem \ref{thm:norm1}. These results thus correspond to opposite regimes of the asymptotic behaviour of $K$ - rapid decay excludes arbitrarily sharp localisation, and asymptotic constancy guarantees it. The intermediate regime, as well as the question of whether definite POVMs arise from physically natural QRFs in QFT, appears to be open.
\end{rem}

\section{Relative Observables}
Let $E$ be a covariant POVM of the form described above, and let $\Ac$ be a von Neumann algebra with a $\sigma$-weakly continuous action $\alpha:
\RR^4\to {\rm Aut}(\Ac)$. For $A\in\Ac$ set
\begin{equation}
    \Y(A)=\int_{\RR^4}\alpha_x(A)\otimes \dd E(x),
\end{equation}
where the integral is understood weakly; the matrix elements
$\ip{\varphi\otimes\xi}{\Y(A)(\psi\otimes\eta)}
=\int_{\RR^4}\ip{\varphi}{\alpha_x(A)\psi}\,\ip{\xi}{E(\dd x)\eta}$
determine $\Y(A)$ uniquely. Existence, normality, complete positivity and
subunitality of $\Y$ are established in \cite{Jamesproj} for an arbitrary von
Neumann algebra $\Ac$ and subnormalised $E$ (the generality needed in QFT,
where local algebras are typically not of type I) provided $E$ extends to
a normal map on $L^\infty(\RR^4)$, as it does here by
Theorem \ref{thm:toller}; explicitly, $\Phi_E(f)=W^*M_fW$ with $M_f$ denoting multiplication by $f$.

Since  $\Y(\1)=\1\otimes E(M)$ and $E(M)\leq \1$, the map $\yen$
 is subunital in general.
For $\varphi,\psi\in\HH$, put $\eta_{\varphi,\psi}=\id_\Ac\otimes\omega_{\varphi,\psi}$, where $\omega_{\varphi,\psi}(T)=\ip{\varphi}{T\psi}$. Then
\begin{equation}
    \eta_{\varphi,\psi}(\Y(A))
    =\int_{\RR^4}\alpha_x(A) \ip{\varphi}{\dd E(x)\psi}
    =\int_{\RR^4}\rho_{\varphi,\psi}(x)\alpha_x(A)\dd^4x.
\end{equation}
Here
\begin{equation}
    \rho_{\varphi,\psi}(x)=\ip{(W\varphi)(x)}{(W\psi)(x)}_{\widetilde{\HH}}
\end{equation}
is an $L^1$-density for arbitrary $\varphi,\psi\in\HH$, and agrees with Toller's smooth density when $\varphi,\psi\in D$. For $\varphi=\psi$ and $\|\psi\|=1$, we write $\rho_\psi=\rho_{\psi,\psi}$; this is a subprobability density, and we refer to $\eta_\psi = \eta_{\psi,\psi}(\Y(A))$ as a restricted relative observable.
\begin{lemma}\label{lem:nocompact}
If $\ip{\psi}{E(M)\psi}>0$, then the essential support of $\rho_\psi$ is noncompact.
\end{lemma}
\begin{proof}
Suppose that $\rho_\psi$ has compact essential support. Then there is a compact set $C\subset M$ such that $\rho_\psi(x)=0$ for almost every $x\notin C$. Since
\begin{equation}
    \rho_\psi(x)=\|(W\psi)(x)\|^2_{\widetilde{\HH}},
\end{equation}
it follows that $(W\psi)(x)=0$ for almost every $x\notin C$. Thus $W\psi$ is a compactly supported element of $L^2(M;\widetilde{\HH})$.
By Paley-Wiener, the Fourier transform of a compactly supported $L^2$-function has an entire $\widetilde{\HH}$-valued extension to $\CC^4$. On the other hand, by the Toller representation,
\begin{equation}
    \widehat{W\psi}(k)=K(k)\psi(k)
\end{equation}
as an element of $L^2(\RR^4;\widetilde{\HH})$, with $K(k)\psi(k)$ extended by zero outside $V^+_{\ac}$. Hence $\widehat{W\psi}$ vanishes on the nonempty open set $\RR^4\setminus {V}^+$. Analyticity forces $\widehat{W\psi}$ to vanish identically, so $W\psi=0$. Consequently
\begin{equation}
    \ip{\psi}{E(M)\psi}=\|W\psi\|^2_{L^2(M;\widetilde{\HH})}=0,
\end{equation}
which proves the result.
\end{proof}
The decay of $\rho_\psi$ at infinity controls the localisation properties of the restricted relative observables. The next proposition gives a simple sufficient condition for rapid decay.
\begin{prop}\label{prop:schwartz}
Let $F_\psi:\RR^4\to\widetilde{\HH}$ be given by $F_\psi(k)=K(k)\psi(k)$ for $k\in V^+_{\ac}$ and by $F_\psi(k)=0$ otherwise. If $F_\psi\in\mathcal S(\RR^4;\widetilde{\HH})$, then $\rho_\psi\in\mathcal S(\RR^4)$.
\end{prop}
\begin{proof}
By Theorem \ref{thm:toller}, $W\psi=\mathcal F^{-1}F_\psi$ as an $L^2$-Fourier transform. Since the inverse Fourier transform is an automorphism of $\mathcal S(\RR^4;\widetilde{\HH})$, one has $W\psi\in\mathcal S(\RR^4;\widetilde{\HH})$. Moreover
\begin{equation}
    \rho_\psi(x)=\|(W\psi)(x)\|^2_{\widetilde{\HH}}.
\end{equation}
For every multi-index $\alpha$,
\begin{equation}
    \partial^\alpha\rho_\psi
    =\sum_{\beta\leq\alpha}c_{\alpha,\beta}
      \ip{\partial^\beta W\psi}{\partial^{\alpha-\beta}W\psi}_{\widetilde{\HH}},
\end{equation}
where $c_{\alpha\,\beta}$ is a multinomial coefficient.
The Schwartz estimates for $W\psi$, together with Cauchy-Schwarz, imply the corresponding estimates for $\rho_\psi$. For example,
\begin{equation}
   \|\rho_\psi\|_{2\gamma,\alpha}\le  
    \sum_{\beta\leq\alpha}c_{\alpha,\beta}
      \|W\psi\|_{\gamma,\beta}\|W\psi\|_{\gamma,\alpha-\beta} ,
\end{equation}
where Schwartz seminorms are defined as usual, $\|f\|_{\beta,\alpha}=\sup_{x\in\RR^4} (1+|x|^\beta)|\partial^\alpha f|$ for $f\in \mathcal S(\RR^4)$, replacing $|\partial^\alpha f|$ by $\|\partial^\alpha f\|_{\widetilde{\HH}}$ for $f\in \mathcal S(\RR^4;\widetilde{\HH})$.
\end{proof}

\begin{defn}\label{def:al}
Let $O\mapsto\mathcal A(O)$ be a local net of von Neumann algebras with quasilocal algebra $\mathcal A$ on Minkowski space $M$. We assume throughout that the local algebras possess a common unit $\id$, (see \cite[p. 849]{haagkastler1964}). An element $B\in\mathcal A$ is \emph{almost local} if, for some $z\in M$ and future-pointing unit timelike vector $u$, there is a family of local operators
\begin{equation}\label{eq:membership}
  B_R\in \Ac(C_R(z,u)),\qquad R>0,
\end{equation}
where $C_R(z,u)= I^+(z-Ru)\cap I^-(z+Ru)$ is the open double-cone with tips at $z\pm Ru$, such that
\begin{equation}\label{eq:decay}
    \sup_{R>0} R^N \|B-B_R\|<\infty \qquad \text{for all $N\in\NN$}
\end{equation}
See \cite{dybalski2014criterion}, and cf. Haag's field version \cite[\S II.4.1]{haag1996}.
\end{defn}
The local approximant $B_R$ belongs to the local algebra of a double-cone about $z$ in the rest frame of $u$. If $B$ is almost local then there is a family of local approximants obeying \eqref{eq:membership} and \eqref{eq:decay}
for every $z\in M$ and future-pointing unit timelike $u$. Note that the double-cones have the property $C_L(z,u)\pm C_{L'}(z',u) = C_{L+L'}(z\pm z',u)$.

\begin{prop}\label{cor:al}
Let $A\in\mathcal A(O)$ be bounded and local, with $O\subset M$ bounded. If $\rho_\psi\in\mathcal S(\RR^4)$, then
\begin{equation}
    B_\psi(A):=\eta_\psi(\Y(A))=\int_{\RR^4}\rho_\psi(x)\alpha_x(A)\dd^4x
\end{equation}
is almost local.
\end{prop}
\begin{proof}
Let $z\in O$, fix any unit future-pointing timelike vector $u$ and suppose that $R$ is sufficiently large that $O\subset C_{R/2}(z,u)$, setting $B_R=0$ for all smaller values of $R>0$. Choose $\chi_R\in C^\infty_c(\RR^4)$ with $0\leq\chi_R\leq1$, $\chi_R(x)=1$ for $x\in C_{R/4}(0,u)$, and $\chi_R(x)=0$ for $x\notin C_{R/2}(0,u)$, and set
\begin{equation}
    B_R=\int_{\RR^4}\chi_R(x)\rho_\psi(x)\alpha_x(A)\,\dd^4x,
\end{equation}
understood in the weak sense.
Since $\alpha_x(A)\in\Ac(O+x)\subset \Ac(C_{R/2}(z+x,u))$ and $\chi_R\rho_\psi$ is supported in $C_{R/2}(0,u)$, isotony gives $B_R\in\Ac(C_R(z,u))$. Moreover,
\begin{equation}
    \|B_\psi(A)-B_R\|\leq \|A\|\int_{\RR^4}|1-\chi_R(x)||\rho_\psi(x)|\,\dd^4x
    \leq \|A\| \int_{\RR^4\setminus C_{R/4}(0,u)} |\rho_\psi(x)|\,\dd^4x.
\end{equation}
Since $\rho_\psi$ is Schwartz, the last integral is $O(R^{-N})$ for every $N$ as $R\to\infty$ (independently of the chosen $\chi_R$).
\end{proof}
 
\begin{lemma}\label{lem:aloccomm}
Let $B_1,B_2$ be almost local and fix any future-pointing timelike unit vector $u$. Then, for every $N\in\NN$, there is $C_N>0$ such that, for all $L>0$ and all $x_1,x_2\in M$ for which
\begin{equation}\label{eq:perp}
    x_1-x_2 \perp C_{2L}(0,u),
\end{equation}
i.e., $x_1-x_2$ belongs to the causal complement of $C_{2L}(0,u)$, one has
\begin{equation}
    \|[\alpha_{x_1}(B_1),\alpha_{x_2}(B_2)]\|\leq C_N L^{-N}.
\end{equation}
In particular, $\|[\alpha_{x_1}(B_1),\alpha_{x_2}(B_2)]\|=O(L^{-\infty})$ as
$L\to\infty$, uniformly in $x_1,x_2$ satisfying \eqref{eq:perp}.
\end{lemma}
\begin{proof}
Fixing any $z\in M$ and any future-pointing unit timelike vector $u$, let $B_{i,R}\in \Ac(C_R(z,u))$ be local approximating families for the $B_i$. Now let $L>0$ and
let $x_1,x_2\in M$ with $x_1-x_2$ spacelike to $C_{2L}(0,u)$. By covariance,
$\alpha_{x_i}(B_{i,L})\in\mathcal A(C_L(z+x_i,u))$, so locality gives
\begin{equation}
    [\alpha_{x_1}(B_{1,L}),\alpha_{x_2}(B_{2,L})]=0,
\end{equation}
because every vector in $C_L(z+x_1,u)-C_L(z+x_2,u)=C_{2L}(x_1-x_2,u) = x_1-x_2-C_{2L}(0,u)$ is spacelike. Hence, writing
\begin{equation}
\begin{aligned}
    [\alpha_{x_1}(B_1),\alpha_{x_2}(B_2)]
    &=[\alpha_{x_1}(B_1-B_{1,L}),\alpha_{x_2}(B_2)]
    +[\alpha_{x_1}(B_{1,L}),\alpha_{x_2}(B_2-B_{2,L})]\\
    &\quad+[\alpha_{x_1}(B_{1,L}),\alpha_{x_2}(B_{2,L})]
\end{aligned}
\end{equation}
and using that $\alpha_{x_1},\alpha_{x_2}$ are isometric, we have
\begin{equation}
\begin{aligned}
    \|[\alpha_{x_1}(B_1),\alpha_{x_2}(B_2)]\|
    &\leq  2\|B_1-B_{1,L}\|\|B_2\| +2\|B_{1,L}\|\|B_2-B_{2,L}\|,
\end{aligned}
\end{equation}
from which, by \eqref{eq:decay} and $\|B_{1,L}\|\le\|B_1\|+\|B_1-B_{1,L}\|$, we arrive at the claim for $L\ge1$; for $L<1$ it holds trivially, since the left-hand side is at most $2\|B_1\|\|B_2\|$.
\end{proof}

\begin{cor}\label{cor:relcomm}
Let $A_i\in\mathcal A(O_i)$ be bounded local observables, with $O_i\subset M$
bounded, and suppose that $\rho_\psi\in\mathcal S(\RR^4)$. Put
\begin{equation}
    B_i=\eta_\psi(\Y(A_i)),\qquad i=1,2.
\end{equation}
Then the conclusion of Lemma \ref{lem:aloccomm} holds for $B_1,B_2$: for every
$N\in\NN$ there is $C_N>0$ such that
\begin{equation}
    \|[\alpha_{x_1}(B_1),\alpha_{x_2}(B_2)]\|\leq C_N L^{-N}
\end{equation}
for all $L>0$ and $x_1,x_2\in M$ satisfying \eqref{eq:perp}.
\end{cor}
\begin{proof}
This follows from Prop. \ref{cor:al} and Lemma \ref{lem:aloccomm}.
\end{proof}

At the formal level, we observe the following. If $\Phi(f)$ is a field, $f\in C^\infty_c(M)$, and $\alpha_x(\Phi(f))=\Phi(f_x)$ with $f_x(y)=f(y-x)$, then
\begin{equation}
    \eta_\psi(\Y(\Phi(f)))=\int_M \rho_\psi(x)\Phi(f_x)\,\dd^4x=\Phi(\rho_\psi*f),
\end{equation}
where
\begin{equation}
    (\rho_\psi*f)(y)=\int_M \rho_\psi(x)f(y-x)\,\dd^4x.
\end{equation}
Thus a compactly supported test function is replaced by a smeared test function. If $\rho_\psi$ is Schwartz, then so is $\rho_\psi*f$, and the resulting field is almost local in Haag's sense \cite{haag1996}. This explains formally why exact microcausality is replaced by rapidly decaying commutators. To avoid domain questions for unbounded fields, we now pass to the Weyl algebra.
For real-valued $f\in C^\infty_c(M)$, let $W(f)$ denote Weyl operators satisfying
\begin{equation}
    W(f)W(g)=\ee^{-\ii E(f,g)/2}W(f+g),
\end{equation}
where $E(f,g)=\int f (E^-g-E^+g)\dvol$ and $E^\pm$ are retarded/advanced Green operators, and suppose translations act by $\alpha_x(W(f))=W(f_x)$. Define
\begin{equation}
    B_f^\psi:=\eta_\psi(\Y(W(f)))=\int_M \rho_\psi(x)W(f_x)\,\dd^4x.
\end{equation}
Then $\|B_f^\psi\|\leq\int_M\rho_\psi(x)\dd^4x\leq1$. The Weyl relations give
\begin{equation}
    [W(f_x),W(g_y)]=-2\ii\sin\left(\frac{1}{2}E(f_x,g_y)\right)W(f_x+g_y),
\end{equation}
and hence
\begin{equation}
    [B_f^\psi,B_g^\psi]
    =-2\ii\int_{M\times M}\rho_\psi(x)\rho_\psi(y)
    \sin\left(\frac{1}{2} E(f_x,g_y)\right)W(f_x+g_y) \,\dd^4x\, \dd^4y.
\end{equation}
Note that $E(f_x,g_y)$ depends only on $y-x$.
  Even if $f$ and $g$ have causally disjoint supports, the integrand will be nonvanishing for suitable
  $x,y$, so microcausality fails.
Let
\begin{equation}
    \Delta_{f,g}:=\{(x,y)\in M\times M:\supp f_x\text{ and }\supp g_y\text{ are causally disjoint}\}.
\end{equation}
On $\Delta_{f,g}$ the integrand vanishes. Therefore
\begin{equation}\label{eq:weylbound}
    \|[B_f^\psi,B_g^\psi]\|
    \leq 2(\rho_\psi\times\rho_\psi)((M\times M)\setminus\Delta_{f,g}).
\end{equation}
This gives a bounded, quantitative version of the preceding formal picture: exact microcausality is replaced by a bound controlled by the probability that two independent frame translations fail to preserve causal disjointness.

\begin{rem}\label{rem:smearedcone}
The bound \eqref{eq:weylbound} entails a statement about the operationally accessible causal structure. For $z\in M$ let
$\chi_{f,g}(z)\in\{0,1\}$ be the set indicator function corresponding to the situation that $(\supp f)+z$ and $\supp g$ fail to be causally disjoint, and let
\begin{equation}
    \kappa_\psi(z):=\int_M \rho_\psi(x)\,\rho_\psi(x-z)\,\dd^4x
\end{equation}
denote the autocorrelation of the localisation density of the frame state $\psi$. Then, for
translated $f_a$, $a\in M$, a change of variables in
\eqref{eq:weylbound} gives
\begin{equation}
    \|[B^\psi_{f_a},B^\psi_{g}]\|\;\leq\;2\,(\kappa_\psi*\chi_{f,g})(a).
\end{equation}
Relative to the frame, the sharp indicator of causal connectedness is replaced by its convolution with $\kappa_\psi$: the boundary of the light
cone is smeared by the localisation uncertainty of the frame state. By Lemma \ref{lem:nocompact}, $\rho_\psi$, and hence $\kappa_\psi$, has non-compact support, so the smearing cannot be removed even in principle: no frame state restores a sharp causal boundary (although for
$\rho_\psi\in\mathcal S(\RR^4)$ there is rapid decay).
\end{rem}

 \section{Commutativity, Compatibility and Disturbance}

A simple way to appreciate the operational content of the microcausality condition in standard
QFT rests on the L\"uders theorem \cite{luders1951uber,busch2009luders}: a nonselective L\"uders measurement of a discrete sharp observable $A=\sum_i \lambda_i P_i$,  resulting in a state update $\rho \mapsto \sum_i P_i \rho P_i$, leaves the statistics of a second sharp observable $B$ undisturbed in every state $\rho$ if and only if $[A,B]=0$. For observables with continuous spectrum, this holds for every discretisation \cite{busch1999unsharp}. For general POVMs commutativity is replaced
by \emph{compatibility}, which implies commutativity only in special cases.

\begin{defn}
    Consider POVMs $A$ and $B$ on measurable spaces $(\Omega_A, \mathcal{F}_A)$ and $(\Omega_B, \mathcal{F}_B)$ respectively (which for the purposes of this paper, we may assume are given as $(\RR^4, \Bor(\RR^4))$). $A$ and $B$ are called \emph{compatible} if there is a \emph{joint} POVM $M$ on
    $\Omega_A \times \Omega_B$ such that for all $X \in \mathcal{F}_A$ and all $Y \in \mathcal{F}_B$, we have $A(X)=M(X \times \Omega_B)$ and $B(Y)=M(\Omega_A \times Y)$.
\end{defn}

This operational definition of compatibility requires the existence of joint probability distributions for $A$ and $B$ in all states, with the distributions for $A$ and $B$ obtained as marginals. If at least one of $A$ and $B$ is projection valued, the compatibility of $A$ and $B$ is equivalent to their commutativity (in the sense that $[A(X),B(Y)]=0$ for all $X$ and $Y$); however, there are many examples of POVMs that are compatible but fail to commute \cite{busch1986,blpy2016}.

For POVMs the relationship between compatibility, (non-)disturbance and commutativity is nuanced.
If some measurement of $A$ leaves $B$ undisturbed, then $A$ and $B$ are compatible (the sequential measurement, $A$ first and then $B$, supplies a joint POVM with marginals $A$ and $B$), but the converse fails, and there are compatible pairs for which every measurement of one POVM disturbs the other \cite{PhysRevA.88.042117}. Compatibility here is therefore not a statement of relativistic causality; the latter depends on the specifics of the available measurements even for compatible observables. The dependence of causal behaviour on how a measurement is implemented, rather
than on the measured observable alone, is already present in ordinary QFT, where microcausality holds exactly. Let $A=\sum_i\lambda_iP_i\in\mathcal{A}(O_1)$ be sharp and local and consider the instrument that measures $A$ and, upon obtaining outcome $\lambda_i$, re-prepares the system in state $\sigma_i$. Here the instrument is then $\mathcal{I}_i(\rho)=\tr[P_i \rho]\sigma_i$, and in the Heisenberg picture its nonselective channel is 
\begin{equation}\label{eq:measure_and_prepare}
\Lambda(B)=\sum_i \tr[\sigma_i B]\,P_i.
\end{equation}
This instrument measures $A$ with the same outcome statistics as the L\"uders instrument, yet $\Lambda(B)\neq B$ for essentially every $B$, independently of the localisation of $B$ --- this measurement signals, even though $A$ commutes with every causally disjoint observable. Indeed, we see that $\Lambda(B)\in \Ac(O_1)$, so $B=\Lambda(B)$ implies $B\in \Ac(O_1)\cap \Ac(O_2)$, which consists only of multiples of the unit under the \emph{extended locality} property that holds for local nets of von Neumann algebras under standard conditions \cite{Landau:1969}.

Causality is therefore not a property of the measured observable, but of the measurement \emph{operation}, and must be formulated in terms of the localisation properties of the operations implementing the measurement. One way to proceed is to assume that the measurement is conducted by a probe QFT locally coupled to the system QFT with the coupled theory also described as a QFT~\cite{fewster2020quantum}, but for the moment we take a more general viewpoint. In Definition~\ref{def:localop}, we give a notion of locality for operations and show that, in the relativised theory, it yields an operational form of microcausality whose accuracy is set by the frame. First, however, we note that relativisation preserves compatibility.

\begin{prop}\label{prop:cpcompat}
    Let $A$ and $B$ be compatible POVMs and let $T: \BB(\HH) \to \BB(\KK)$ be a normal unital completely positive map. Then $T\circ A$ and $T\circ B$ are compatible.
\end{prop}
The proof follows immediately from noting that if $M$ is a joint POVM for $A$ and $B$, then $T\circ M$ is a joint POVM for $T\circ A$ and $T\circ B$ \cite{heinosaari2015incompatibility}. The same conclusion holds for subnormalised POVMs and subunital normal completely positive maps, after making the appropriate changes.

Applying the completely positive map $T_\psi=\eta_\psi\circ\Y$ to a joint PVM makes the contrast between commutativity and compatibility explicit. For causally disjoint $f$ and $g$ the Weyl operators $W(f)$ and $W(g)$ commute, and there is a joint PVM $P$ whose marginals are the spectral measures $P_f$ and $P_g$ of $W(f)$ and $W(g)$. By Proposition \ref{prop:cpcompat}, $T_\psi\circ P$ is a joint POVM for the compatible pair $T_\psi\circ P_f$ and $T_\psi\circ P_g$, whose first moments are the non-commuting operators $B_f^\psi$ and $B_g^\psi$, as seen by the calculation
\begin{equation}
    \int_\CC \lambda \,\dd (T_\psi\circ P_f)(\lambda) =T_\psi \left(\int_\CC \lambda \,\dd  P_f(\lambda) \right) =T_\psi(W(f)) = B_f^\psi.
\end{equation}
 Thus relativisation (and restriction) may destroy commutativity, but nonetheless preserves compatibility.

We now turn to disturbance. Signalling concerns the nonselective channel defined by an instrument, and so we work directly at the level of channels. In addition to the localisation of observables, the localisation of operators is also part of the philosophy of local quantum physics;  operations implemented by elements of a local algebra appear already in Haag and Kastler \cite{haagkastler1964}. That the locality of an observable does not by itself ensure the causal behaviour
of its measurement was stressed by Sorkin \cite{sorkin1993impossible} (see also \cite{BorstenJubbKells2021}), and the identification of operations that are both causal and physically realisable has since been pursued from several directions: measurement schemes in which the system is coupled locally to a
probe \cite{fewster2020quantum,BostelmannFewsterRuep2021}; the direct analysis of state-update rules for scalar fields \cite{jubb2022};
the construction of regularised, non-signalling measurements of field
operators and their spacetime-extended
generalisations \cite{Oeckl2025,Oeckl2026,OecklZampeli2026}; and the general study of causal and realisable local operations \cite{mandrysch2026causality}.

\begin{defn}\label{def:localop}
By an \emph{operation} we mean a normal unital completely positive map $\Lambda:\BB(\HH)\to\BB(\HH)$ (a channel, in the Heisenberg picture). Every operation admits a Kraus decomposition
\begin{equation}
    \Lambda(B)=\sum_{i\geq0}K_i^*\,B\,K_i,\qquad \sum_{i\geq0}K_i^*K_i=\id,
\end{equation}
with both sums converging $\sigma$-weakly \cite{kraus1983}. An operation $\Lambda$ is \emph{local in a bounded region} $O\subset M$ if it admits such a decomposition with $K_i\in\mathcal{A}(O)$ for every $i$, and \emph{almost local} if it admits one in which every $K_i$ is almost local. The Kraus decomposition of a given operation is not unique; locality and almost locality require one decomposition of the stated form; Remark \ref{rem:rigidity} below pertains to every decomposition.
\end{defn}
We caution immediately that `locality' of an operation does not imply that it can be realised by local physical interactions within the given region. More comments on this appear below. (Note also that our terminology here differs from that of~\cite{mandrysch2026causality}, in which an instrument is said to be local to a compact set $K$ if its measurement channel acts trivally on the algebra associated with the causal complement of $K$; see however Sec.~IV of~\cite{mandrysch2026causality} for situations in which their definition implies that the instrument is local, in our sense, to any region containing $K$.)
The identity channel (a single Kraus operator $K_0=\id$) is local in every region. The L\"uders channel of a discrete POVM in $\mathcal{A}(O_1)$ --- effects $A_1,\dots,A_n\in\mathcal{A}(O_1)$ with $\sum_{i=1}^n A_i\leq\id$, completed by $A_0:=\id-\sum_{i=1}^n A_i$ --- is local in $O_1$: its Kraus operators are $A_i^{1/2}\in\mathcal{A}(O_1)$ for $0\leq i\leq n$, since the continuous functional calculus preserves $\mathcal{A}(O_1)$. Thus the L\"uders channel of every discrete local POVM is local in the sense of
Definition \ref{def:localop}. The measure-and-prepare channel  of~\eqref{eq:measure_and_prepare} is, by contrast, local in no bounded region, as part (i) of the following proposition shows.

\begin{prop}\label{prop:nosignal}
Let $\Lambda$ be an operation with Kraus operators $K_i$ as in Definition \ref{def:localop}.
\begin{enumerate}
    \item[(i)] If $\Lambda$ is local in $O_1$ and $O_2$ is causally disjoint from $O_1$, then $\Lambda(B)=B$ for every $B\in\mathcal{A}(O_2)$.
    \item[(ii)] Suppose $\Lambda$ is almost local with a finite Kraus decomposition
$K_0,\dots,K_n$, let $B$ be almost local, fix a future-pointing unit timelike
vector $u$, and for $x\in M$ let $\Lambda_x:=\alpha_x\circ\Lambda\circ\alpha_x^{-1}$
denote the translated operation. Then, for every $N\in\mathbb{N}$, there is $C_N>0$ such that, for all $L>0$ and all $x_1,x_2\in M$ satisfying \eqref{eq:perp},
\begin{equation}\label{eq:nosignal}
\|\Lambda_{x_1}(\alpha_{x_2}(B))-\alpha_{x_2}(B)\|\le C_N L^{-N}.
\end{equation}
\end{enumerate}
\end{prop}

\begin{proof}
Since $\sum_{i}K_i^*K_i=\id$, with all sums converging $\sigma$-weakly,
\begin{equation}
    \Lambda(B')-B'
    =\sum_{i}K_i^*\big[B',K_i\big].
\end{equation}
For (i), take $B'=B\in\mathcal{A}(O_2)$; each $K_i\in\mathcal{A}(O_1)$, so every term vanishes by microcausality. For (ii), the decomposition is finite and $\|K_i\|\leq1$, so, using that $\Lambda_{x_1}$ has Kraus operators $\alpha_{x_1}(K_i)$,
\begin{equation}
    \|\Lambda_{x_1}(\alpha_{x_2}(B))-\alpha_{x_2}(B)\|
    \leq\sum_{i=0}^{n}\big\|[\alpha_{x_2}(B),\alpha_{x_1}(K_i)]\big\|,
\end{equation}
and Lemma \ref{lem:aloccomm} bounds each term by $C_N' L^{-N}$, uniformly in $x_1,x_2$ as in the statement.
\end{proof}

\begin{rem}\label{rem:rigidity}
Part (i) admits a converse, showing that the notion of locality in Definition \ref{def:localop} is forced by causality. Suppose $\Lambda(B)=B$ for every $B\in\mathcal{A}(O_2)$ and every bounded $O_2$ causally disjoint from $O_1$. For any Kraus decomposition of $\Lambda$ and any such $B$,
\begin{equation}
    \sum_{i}\big(K_iB-BK_i\big)^*\big(K_iB-BK_i\big)
    =\Lambda(B^*B)-\Lambda(B^*)B-B^*\Lambda(B)+B^*B=0,
\end{equation}
since $B^*$ and $B^*B$ also lie in $\mathcal{A}(O_2)$; a vanishing sum of positive terms, whence $[K_i,B]=0$ for every $i$
\cite{choi1974,arias2002}. Thus each $K_i$ lies in the commutant of the von Neumann algebra $\mathcal{A}(O_1')$ generated by the local algebras $\mathcal{A}(O_2)$ of bounded regions $O_2$ causally disjoint from $O_1$. Where Haag duality holds for $O_1$ (for instance, when $O_1$ is a double cone in a theory satisfying Haag duality), $\mathcal{A}(O_1')'=\mathcal{A}(O_1)$, so
every Kraus operator of $\Lambda$ lies in $\mathcal{A}(O_1)$: the operations that cannot signal out of $O_1$ are precisely those local in $O_1$. This observation appears as Proposition 49 of \cite{mandrysch2026causality}; the underlying fact - that a channel acting as the identity on a $^*$-algebra has Kraus operators in its commutant, is due to Lindblad \cite{lindblad1999general}, whose proof rests on the equality situation of the Schwarz inequality for completely positive maps \cite{choi1974}. The same identity, applied to the spectral projections of a sharp observable, yields the nontrivial part of the L\"uders theorem.
\end{rem}

We note that Proposition \ref{prop:nosignal} and Remark \ref{rem:rigidity} concern two regions:as Sorkin observed \cite{sorkin1993impossible} (see \cite[Sec. II.E]{mandrysch2026causality} for a formulation in terms of channels), locality of the Kraus operators alone does not preclude superluminal signalling in protocols involving three or more regions, and the characterisation of the operations that are both physically realisable and fully causal is delicate \cite{fewster2020quantum,jubb2022,mandrysch2026causality}; these matters lie beyond the scope of this paper,  but it is worth emphasising that the same issues also arise in classical field theories \cite{MuchVerch:2023}.  

To bring measurements of relativised observables within the scope of part (ii), it remains to verify that the Kraus operators of their L\"uders channels are again almost local.

\begin{lemma}\label{lem:sqrtal}
Let $E$ be an almost local effect. Then $E^{1/2}$ and $(\mathrm{id}-E)^{1/2}$
are almost local.
\end{lemma}

\begin{proof}
Fix $z\in M$ and a future-pointing unit timelike vector $u$, and let
$E_R\in\mathcal{A}(C_R(z,u))$ be local approximants, which may be taken self-adjoint, with $\varepsilon_R:=\|E-E_R\|=O(R^{-\infty})$. Let $f(t)=\min\{\max\{t,0\},1\}$; then $F_R:=f(E_R)\in\mathcal A(C_R(z,u))$ are effects with
$\|E_R-F_R\|\leq\varepsilon_R$ by the functional calculus, so
$\|E-F_R\|\leq 2\varepsilon_R$. Since the square root is monotone and vanishes at $0$, $\|E^{1/2}-F_R^{1/2}\|\leq\|E-F_R\|^{1/2}$ \cite[Thm. X.1.1]{bhatia1997matrix} (the proof there is dimension-independent, using only operator monotonicity of the square root), which is $O(R^{-\infty})$; as $F_R^{1/2}\in\mathcal{A}(C_R(z,u))$, $E^{1/2}$ is almost local. Similarly,
$\|(\id-E)^{1/2}-(\id-F_R)^{1/2}\|\leq\|E-F_R\|^{1/2}$ and
$(\id-F_R)^{1/2}\in \mathcal{A}(C_R(z,u))$, giving the second claim.
\end{proof}

\begin{cor}\label{cor:relnosignal}
Suppose $\rho_\psi\in\mathcal S(\RR^4)$, let $A_1,\dots,A_n\in\mathcal{A}(O_1)$ be effects with $\sum_{i=1}^n A_i\leq\id$, and put $E_i=\eta_\psi(\Y(A_i))$ and $E_0:=\id-\sum_{i=1}^n E_i$. Then the nonselective L\"uders channel of the relativised observable $\{E_i\}_{i=0}^n$,
\begin{equation}
    \Lambda(B')=\sum_{i=0}^{n}E_i^{1/2}\,B'\,E_i^{1/2},
\end{equation}
is an almost local operation, and its translate $\Lambda_{x_1}$ is the L\"uders channel of the translated observable $\{\alpha_{x_1}(E_i)\}_{i=0}^n$. Consequently, by Proposition \ref{prop:nosignal}(ii), a nonselective L\"uders measurement of $\{E_i\}$ alters the expectation of any relativised local observable at causal distance $L$ in the sense of \eqref{eq:perp} by at most $O(L^{-\infty})$.
\end{cor}

\begin{proof}
Since $\Y$ is completely positive and subunital and $\eta_\psi$ is unital and completely positive, each $E_i$ ($1\le i\le n$) is an effect, almost local by Proposition \ref{cor:al}, and
hence so is $E:=\sum_{i=1}^n E_i=\eta_\psi(\Y(\sum_iA_i))\le\mathrm{id}$, a sum
of approximants for the $E_i$ (taken with a common $z$ and $u$) serving as
approximants for $E$. By Lemma \ref{lem:sqrtal}, the Kraus operators $E_i^{1/2}$
($1\le i\le n$) and $E_0^{1/2}=(\mathrm{id}-E)^{1/2}$ are almost local. Hence
$\Lambda$ is almost local, and since $\alpha_{x_1}$ commutes with the
functional calculus, $\Lambda_{x_1}$ is as claimed. Proposition \ref{prop:nosignal}(ii) applies.
\end{proof}

Proposition \ref{prop:nosignal} and Corollary \ref{cor:relnosignal} together provide what is left of of microcausality in the relativised theory: no operation with a finite almost local Kraus decomposition, including the L\"uders measurements of every finite-outcome relativised local observable, can affect the expectation of a distant relativised observable by more than $O(L^{-\infty})$. We stress that \eqref{eq:nosignal} should not be read as asserting the possibility of superluminal signalling. By Proposition \ref{prop:toller1} and Lemma \ref{lem:nocompact}, relativised observables admit no sharp localisation in bounded regions, so that the spacetime labels appearing in any relative no-signalling statement are themselves defined only to the accuracy set by the frame state.

Thus far we have considered how the local and causal properties of an observable fare under relativisation. We now turn to the question of
the physical requirements of the frame for the relative and absolute descriptions to agree closely. Relative to a frame state $\psi$,
the actually realisable counterpart of a system effect $A$ is the restricted relative observable $T_\psi(A)=\eta_\psi(\Y(A))$, and the quality of approximation is given by $D=\|T_\psi(A)-A\|$. In the quantum mechanical setting, bounds on $D$ were obtained in \cite{miyadera2016}, where good approximation was shown to
require a broad distribution of the symmetry generator in the frame state.
For translations in Minkowski space the relevant property of the frame is its ability to resolve a displacement $y$, quantified by the overlap $c_\psi(y)=(\psi,U(y)\psi)$ of the frame state with its translate: $|c_\psi(y)|$ close to $1$ means that the frame cannot resolve $y$. The following proposition shows that if $A$ is sensitive to the displacement $y$,
then accurate recovery of $A$ forces $|c_\psi(y)|$ to be small. The bound also involves the unsharpness $V(A)=\|A-A^2\|$ of $A$, which vanishes exactly when $A$ is a projection.

\begin{prop}\label{prop:tradeoff}
Let $A$ be a system effect, suppose the action $\alpha$ is implemented by a strongly continuous unitary representation $U_S$ of $\RR^4$, and put
\begin{equation}
    T_\psi(A)=\eta_\psi(\Y(A)),\qquad D=\|T_\psi(A)-A\|,
    \qquad V(A)=\|A-A^2\|.
\end{equation}
For $y\in M$, let $c_\psi(y)=\ip{\psi}{U(y)\psi}$. If $c_\psi(y)\neq0$, then
\begin{equation}\label{eq:tradeoff}
    \|\alpha_y(A)-A\|\leq 2D+
    \frac{2\sqrt{1-|c_\psi(y)|^2}}{|c_\psi(y)|}\sqrt{2D+V(A)}.
\end{equation}
\end{prop}
\begin{proof}
Set $X=\Y(A)$ and $\eta=\eta_\psi$. Since $0\leq A\leq \id$ and $\Y$ is completely positive and subunital, $X$ is an effect. The  map $\eta$ is unital and completely positive, so the estimate given in \cite[eq. (9)]{miyadera2016} gives
\begin{equation}
    \|\eta(X^2)-\eta(X)^2\|\leq 2D+V(A).
\end{equation}
Moreover $X$ is invariant, so $[X,U_S(y)\otimes U(y)]=0$, while
\begin{equation}
    \eta(U_S(y)\otimes U(y))=c_\psi(y)U_S(y).
\end{equation}
The Cauchy-Schwarz inequality for completely positive maps \cite[Lem. 1]{janssens2017unifying} gives
\begin{equation}
    \|[T_\psi(A),c_\psi(y)U_S(y)]\|
    \leq 2\sqrt{1-|c_\psi(y)|^2}\sqrt{2D+V(A)}.
\end{equation}
Dividing by $|c_\psi(y)|$ and using $\|[A,U_S(y)]\|\leq \|[T_\psi(A),U_S(y)]\|+2D$ gives \eqref{eq:tradeoff}, up to the sign convention in the implementation of translations.
\end{proof}
\begin{rem}\label{rem:tradeoff}
 If $A$ varies appreciably under a translation $y$, then accurate recovery of $A$ by relativisation followed by restriction requires $D$ to be small and hence the frame state to have sufficiently small $|c_\psi(y)|=|\ip{\psi}{U(y)\psi}|$. In other words, and bearing in mind that $c_\psi(0)=1$, the frame must itself resolve the displacement $y$. Since $c_\psi$ is related to the QRF energy-momentum distribution, resolving shorter displacement scales requires broader energy-momentum support.
\end{rem}

To quantify this last point slightly more, we argue as follows.
Recalling that $c_\psi(y)=\ip{\psi}{U(y)\psi}$, we note that
\begin{equation}
    |c_\psi(y)| \ge \Re c_\psi(y) = \int \cos(p\cdot y) \,\dd\mu_\psi(p)
    \ge (1-M_{\psi,\theta})\cos(\theta) - M_{\psi,\theta}
\end{equation}
holds for any $0<\theta<\pi/2$, where $M_{\psi,\theta} = \mu_\psi(\{p\in\overline{V}^+: |p\cdot y|> \theta\})$, which follows from the rather crude bound that $\cos (p\cdot y)\ge \cos\theta$ for $|p\cdot y|\le\theta$ and $\cos (p\cdot y)\ge -1$ otherwise.
Supposing that $|c_\psi(y)|<\epsilon$ and rearranging,
\begin{equation}
    M_{\psi,\theta} \ge \frac{\cos(\theta) - \epsilon}{1+\cos(\theta)}.
\end{equation}
If $\theta$ is chosen small but nonzero, and $\epsilon\ll 1$, then $M_{\psi,\theta}\gtrsim \tfrac{1}{2}$. Thus, a significant proportion of the energy-momentum distribution of $\psi$ lies in the set $\{p\in\overline{V}^+: |p\cdot y|> \theta\}$. If, for example, $y$ is a timelike vector, then all momenta $p\in\overline{V}^+$ with
$|p\cdot y|> \theta$ have energy greater than $\theta/\sqrt{y\cdot y}$ in the rest-frame of $y$. Alternatively, if $y$ is spacelike and $|p\cdot y|>\theta$ then $p$ has energy greater than $\theta/\sqrt{-y\cdot y}$ in any frame for which $y$ is purely spatial, i.e., orthogonal to the observer's $4$-velocity. To see this, let $u$ be the $4$-velocity, then $p\in \overline{V}^+$ implies $(p\cdot u)^2\ge (p\cdot y)^2/(-y\cdot y)\ge \theta^2/(-y\cdot y)$. In either case, resolving short spacetime displacements requires high energies in an associated frame of reference.

\section{Concluding Remarks}
The spectrum condition obstructs the existence of ideal reference frames for spacetime translations in QFT: translation-covariant POVMs are never projection-valued, and they fail the norm-$1$ property in many cases. Relative to such a frame, local observables become almost local, exact microcausality is replaced by rapid commutator decay controlled by the localisation density of the frame state. From an  operational standpoint, in which spacetime structure is abstracted from relations among observables, the results above indicate that the causal structure relative to any realistic frame is intrinsically smeared (Remark \ref{rem:smearedcone}), and the sharp light cone appears as an idealised limit of frame localisation - a limit which, by Corollary \ref{cor:nopvm}, no physical frame satisfies. We
leave the systematic development of this perspective, and the associated question of the emergence of classical causal structure, to future work.

Several further questions remain open. The frames with unbounded $h$, the only remaining candidates for the norm-$1$ property (Remark \ref{rem:Tol5}), need a physical characterisation. Relativised two-point functions of Hadamard states may provide a natural class of examples. Whether the operational compatibility preserved under relativisation excludes obstructions to measurements of the kind identified by Sorkin \cite{sorkin1993impossible} is open, and connects with the measurement framework of \cite{fewster2020quantum}. The relation between causal and realisable operations is delicate already in the absolute setting \cite{mandrysch2026causality} and we leave the relational setting for the future. Finally, we note that the necessary unsharpness of time observables may play a role in the ubiquity of the phenomenon of type change \cite{chandrasekaran2023algebra,fewster2025quantum,de2025gravitational,de2025crossed}.

{\bf Note added}. During the completion of this work, we became aware of the work of Agullo and Mele \cite{Fabio}, who investigate QRFs also in a QFT setting with  Gaussian-modulated POVMs as frame observables, which do not satisfy the norm-$1$/localisability property. We look forward to comparing our findings to theirs, particularly in connection with Corollaries \ref{cor:HS_norm1} and \ref{cor:sem1}. We also became aware of Freidel and Kirklin \cite{freidel2026gravitational}, in which the  quantum reference frame for reparametrisations of a gravitational null ray
is necessarily non-ideal; positivity of the null energy excludes a sharp frame observable. Though the context is different, this gives further indication that unsharp QRFs are likely to be typical in relativistic quantum settings quite generally.

\begin{acknowledgments}
We thank Daan Janssen, Kasia Rejzner, Jan G\l owacki and Samuel Fedida for useful discussions. We also thank Laurent Freidel for pointing out \cite{freidel2026gravitational}. The work of CJF is supported in part by EPSRC Grant EP/Y000099/1 to the University of York. 
For the purpose of open access, the authors have applied a creative commons attribution (CC BY) licence to any author accepted manuscript version arising. 
\end{acknowledgments}

\makeatletter\@booleantrue\longbibliography@sw\makeatother

\bibliography{bib.bib}
\end{document}